\documentclass[aps,pra,groupedaddress]{revtex4}
\usepackage{amsmath}
\usepackage{amsfonts}
\usepackage{amssymb}
\usepackage{epsf}
\newtheorem{The}{Theorem}
\newtheorem{Def}[The]{Definiton}
\newtheorem{Lemma}[The]{Lemma}
\newtheorem{Pro}[The]{Proposition}

\newenvironment{proof}{\par\noindent\textit{Proof.\ }}{\hfill $\square$ \vspace{1em}}

\def\Cx{\mathbb C}
\def\idty{\openone}

\def\norm#1{\|#1\|}
\def\ketbra#1#2{\vert#1\rangle\langle#2\vert}

\newcommand{\tr}{\operatorname{tr}}
\def\id{{\rm id}}

\def\HH{{\cal H}}\def\KK{{\cal K}}\def\BB{{\cal L}}
\newcommand{\e}{\mathrm{e}}
\newcommand{\im}{\mathrm{i}}
\newcommand{\dd}{\mathrm{d}}

\def\Tcorr{T_{\rm corr}}

\def\SU#1{{\rm SU}_{#1}}
\newcommand{\fid}{\mathcal{F}}
\newcommand{\range}{\operatorname{range}}
\newcommand{\rin}{\tilde{R}_\alpha}

\begin{document}
\title{Quantum Lost and Found}
\author{M. Gregoratti}
\email{gregoratti@mate.polimi.it}
\affiliation{Dip.\ Mat., Politecnico di Milano, piazza Leonardo da Vinci 32, I-20133 Milano, Italy}
\author{R.~F. Werner} \email{r.werner@tu-bs.de} 
\affiliation{Inst.\ Math.\ Phys., TU-Braunschweig, 
Mendelssohnstra{\ss}e 3, D-38106 Braunschweig, Germany} 
\begin{abstract} We consider the problem of correcting the errors 
incurred from sending classical or quantum information through a 
noisy quantum environment by schemes using classical information 
obtained from a measurement on the environment. We give a 
conditions for quantum or classical information (prepared in a 
specified input basis B) to be corrigible based on a measurement 
M. Based on these criteria we give examples of noisy channels such 
that (1) no information can be corrected by such a scheme (2) for 
some basis B there is a correcting measurement M (3) for all bases 
B there is an M (4) there is a measurement M which allows perfect 
correction for all bases B. The last case is equivalent to the 
possibility of correcting quantum information, and turns out to be 
equivalent to the channel allowing a representation as a convex 
combination of isometric channels. Such channels are doubly 
stochastic but not conversely. \end{abstract} \maketitle 
\section{Introduction}
A fundamental feature of open quantum systems \cite{Davies} is the 
possibility of irreversible evolutions, which arise from coupling 
the system under consideration to an environment, a unitary 
evolution of the system together with its environment, and a 
subsequent reduction to the system. In fact, this mechanism for 
obtaining decoherence is universal, because any irreversible 
evolution can be obtained from such a scheme. In this sense 
decoherence always arises from the loss of information to the 
environment. 

Decoherence is, of course, one of the main problems encountered in 
the realization of quantum information tasks, from quantum 
cryptography all the way to the quantum computer. It is therefore 
natural to try to combat decoherence not just by the usual error 
correcting codes \cite{Nielsen,QECC} at system level, but by going 
to the environment, retrieving some of the lost information, and 
using it to get a more efficient error correction (see for example 
the ``embedded quantum codes'' of Alber et.~al~\cite{Alber}). One 
problem with this idea is immediately obvious: in many cases the 
``environment'' is just too big to be controlled sufficiently well 
to make the necessary measurements. For example, the correction of 
errors due to spontaneous emission \cite{Plenio} would require to 
catch all the emitted photons, which seems impossible in many 
experimental setups. On the other hand, there are experiments 
(e.g., inside a cavity) where this may become feasible. What a 
theoretical analysis in abstract terms can do here is to {\it tell 
in advance if there is a chance}. That is, if the noisy channel 
describing the effective evolution of the system is known, can we 
decide if there exists a suitable measurement on the environment, 
which finds just the (classical) information needed to restore the 
system? 

It turns out that channels differ very much in this respect, and 
that the differences are not simply a question of more or less 
``noise''. In fact, the noisiest channel of all, the completely 
depolarizing channel, may be corrected perfectly with such a 
scheme. On the other hand, there are channels not allowing such 
correction, even if one only wants to transmit classical 
information. 

Our paper is organized as follows:
 In Section~\ref{sec:general}, we set up the framework and the
basic correction scheme, and we prove the basic criteria for the 
existence of a correction scheme for a given channel, and a given 
measurement on the environment. In Section~\ref{sec:types} we 
describe the degrees of corrigibility, which are then shown by 
explicit examples to be distinct in Section~\ref{sec:examples}. It 
turns out that qubit channels are more easily shown to be 
corrigible, which will be shown in Section~\ref{sec:qubits}. In 
real life applications, it will often not be possible to achieve a 
complete correction of errors. Instead one is interested in the 
{\it optimal} correction scheme. One part of this optimization 
problem is solved in Section~\ref{OptRec}: for fixed measurement 
on the environment we determine the optimal recovery operation. In 
the main paper the environment is assumed to be initially in a 
pure state. Without that assumption corrigibility may be 
dramatically worse, even for qubits (Section~\ref{mixedenv}). In 
the final Section~\ref{sec:locc} we remark on an alternative 
correction scheme, in which there is more classical communication 
between the system and the `Lost and Found Office' in the 
environment: if we allow rounds of classical communication and 
local quantum operations, correction may become possible in some 
situations, in which our scheme would not work.

  \par\vskip40pt

\section{General Results}\label{sec:general}
\subsection{The correction scheme}\label{CorrSch}

In this Section we establish some notation, and describe the 
general scheme (see Fig.~1) for correcting information using 
information from the environment. Throughout we will consider a 
channel $T$ transforming systems with Hilbert space $\HH_1$ into 
systems with a possibly different Hilbert space $\HH_2$. Both 
Hilbert spaces are assumed to be finite dimensional. We work in 
the Schr\"odinger picture, so the channel is given by a map taking 
each input density operator $\rho$ on $\HH_1$ to a density 
operator $T(\rho)$ on $\HH_2$. This map extends to a linear 
mapping $T:\BB(\HH_1)\to\BB(\HH_2)$, where $\BB(\HH)$ denotes the 
space of all linear operators on a Hilbert space $\HH$. Of course, 
$T$ must be completely positive and trace preserving 
\cite{Nielsen}, which is equivalent to the possibility of 
obtaining it by unitary coupling to an environment. It is natural 
to distinguish also the Hilbert spaces $\KK_1$ and $\KK_2$ of the 
environment before and after the coupling. Then the interaction is 
given by a unitary operator 
$U:\HH_1\otimes\KK_1\to\HH_2\otimes\KK_2$. If the initial states 
of system and environment are density operators $\rho$ and 
$\rho_0$, respectively, an observable $A\in\BB(\HH_2)$ measured on 
the system after the interaction has expectation 
\begin{eqnarray}\label{TfromU} 
  \tr(T(\rho) A)
    &=& \tr\Bigl(U(\rho\otimes\rho_0)U^*(A\otimes\idty)\Bigr)\;,
\quad\mbox{i.e.}\nonumber\\
  T(\rho)&=&\tr_{\KK_2}\Bigl(U(\rho\otimes\rho_0)U^*\Bigr)\;,
\end{eqnarray} where the trace in the second line is the partial 
trace over the Hilbert space specified. In this expression we have 
simply discarded the environment by choosing an observable of the 
form $(A\otimes\idty)$ on the combined system. A  measurement on 
the environment $\KK_2$ would be given by a family of operators 
$M_\alpha\in\BB(\KK_2)$, indexed by the classical outcomes 
``$\alpha$'', with $M_\alpha\geq0$ and $\sum_\alpha 
M_\alpha=\idty$. Inserting $M_\alpha$ instead of $\idty$ into 
(\ref{TfromU}) we get \begin{equation}\label{TafromU} 
  \tr(T_\alpha(\rho) A)
    = \tr\Bigl(U(\rho\otimes\rho_0)U^*(A\otimes M_\alpha)\Bigr)\;,
\end{equation}
where the completely positive map $T_\alpha$ is a selective channel giving the (non-normalized)
output state $T_\alpha(\rho)$ of the subensemble of systems for which the measurement performed
after the interaction on the environment has given the result ``$\alpha$'':
\begin{figure}
   \epsfxsize=8cm \epsfbox{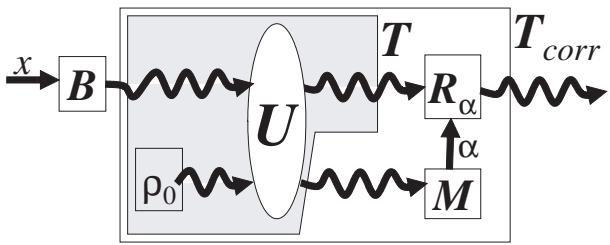}
   \caption{The basic correction scheme. The noisy channel $T$ is represented by
   the shaded shape, and consists of a unitary coupling $U$ of the system to an environment
   in state $\rho_0$. The result $\alpha$ of the measurement $M$ on the environment is
   used to select the recovery operation $R_\alpha$, resulting in the overall corrected
   channel $\Tcorr$. }
\end{figure}
 the probability of getting $\alpha$ is $\tr T_\alpha(\rho)$, the normalized output state for the
corresponding subensemble is $T_\alpha(\rho)/\tr T_\alpha(\rho)$, while $\tr(T_\alpha(\rho) A)/\tr
T_\alpha(\rho)$ is the expectation of $A$ in that subensemble. The structure of a channel,
decomposed into a sum $T=\sum_\alpha T_\alpha$ of selective operations, is called an {\it
instrument} \cite{Davies}. It is the general form of a quantum operation yielding classical
information (i.e., $\alpha$) together with a quantum output.

We can now introduce the idea of correction:  this will be a 
restoring operation taking systems with Hilbert space $\HH_2$ to 
systems with the same Hilbert space $\HH_1$ as the inputs. We 
allow this operation to depend in an arbitrary way on the 
available classical information $\alpha$. Hence it is given by a 
family of channels $R_\alpha:\BB(\HH_2)\to\BB(\HH_1)$. After 
correction, the state of the subensemble for which  the 
measurement has given the  result $\alpha$, will be 
$R_\alpha(T_\alpha(\rho))$ up to the normalization factor 
$\tr(T_\alpha(\rho))$. The overall corrected channel is built from 
these conditional operations by ignoring the intermediate 
information $\alpha$, and is the sum of these contributions: 
\begin{equation}\label{Tcorr} 
  \Tcorr=\sum_\alpha  R_\alpha\circ T_\alpha:\BB(\HH_1)\to\BB(\HH_1)\;,
\end{equation}
where $\circ$ means composition of maps. We will say that the scheme {\it restores
quantum information}, if the corrected channel is the ideal channel on $\HH_1$, i.e.,
\begin{equation}\nonumber
    \Tcorr=\id_1\;.
\end{equation}
By the ``No information without perturbation'' Theorem, this condition is equivalent to
$R_\alpha\circ T_\alpha = c_\alpha \id_1$, for some $c_\alpha\geq0$, $\sum_\alpha c_\alpha=1$. This
implies firstly that we get the result $\alpha$ with probability
$\tr(T_\alpha(\rho))=\tr\bigl(R_\alpha(T_\alpha(\rho))\bigr)=c_\alpha$, which is independent of
$\rho$. That is, the results $\alpha$ contain no information whatsoever about the initial state.
Moreover, this special form implies that the original quantum state is restored, not only for the
whole ensemble, but also for each subensemble selected according to the result $\alpha$.

We will also consider in detail the question if classical information can be transmitted
faithfully. In this case we use a particular basis $B$ in the input space $\HH_1$ to encode the
information. Let $B_x$, $x=1,\ldots,\dim\HH_1$, denote the one-dimensional projections onto the
basis vectors. Then we will say that the scheme {\it restores classical information in basis $B$},
if
\begin{equation}\label{correctclass}
  \Tcorr(B_x) =B_x\;.
\end{equation}
Now condition \eqref{correctclass} is equivalent to $R_\alpha\circ T_\alpha(B_x) = B_x \cdot
\tr T_\alpha(\rho)$ and again we can say that also for each $\alpha$-subensemble classical information
is restored.

\subsection{Independence of Coupling}
Whether or not we can find a correction scheme in principle depends not just on the noisy channel
$T$, but on the particular way in which it is realized by coupling, i.e., on the unitary
interaction $U$ and the initial state $\rho_0$ of the environment. However, if $\rho_0$ happens to
be a pure state this dependence simplifies considerably. It is clear from (\ref{Tcorr}) that
environments in which the same decompositions $T=\sum_\alpha T_\alpha$ of a given channel $T$ can
be realized are completely equivalent in terms of retrieval of lost information in our scheme. And it is
equally clear that, if we can further decompose some $T_\alpha=T_{\alpha'}+T_{\alpha''}$, we strictly
improve our chances for correction: we can always choose the same recovery operation
$R_{\alpha'}=R_{\alpha''}=R_\alpha$ reproducing the previous result, but it might be helpful to
take $R_{\alpha'}\neq R_{\alpha''}$ instead. In
this sense the following Theorem tells us that all realizations of a channel $T$ by coupling to an
{\it initially pure} environment are equivalent and that for all of them there are no limitations in the
choice of the decomposition.

Assuming a pure environment may seem counterintuitive, if the environment is seen as that ``large
and uncontrolled system out there''. However, in situations in which our correction scheme might be
applied, a good control of the environment is needed anyhow. The ``environment'' is then only that
part with which the system is in contact, and which together with the system is sufficiently well
isolated from the rest of the world. As the following Theorem and Section~\ref{mixedenv} show, if we
want to ``tell in advance if there is a chance'', without using information about the details of the
coupling, we should make this assumption in any case.

Pure environments are also in keeping with the general idea that all decoherence arises from
coupling to an environment. Only with a pure environment we can separate the combined effects of
interaction and tracing out on the one hand, from the transfer of impure information from the
environment to the system on the other. Not that this is an absolute distinction: by purifying the
environment, i.e., by considering it as a subsystem of a larger system in a pure state, we can
always represent the `transfer of impurity' as a dynamical coupling to one part of a pure entangled
system.

Before stating and proving the Theorem, let us recall some basic 
facts about possible decompositions and refinements of a given 
channel $T$. The finest, i.e., the most informative measurements 
are those for which the decomposition $T=\sum_\alpha T_\alpha$ 
allows no proper refinement. This means that the only further 
decompositions of $T_\alpha=T_{\alpha'}+T_{\alpha''}$ into 
completely positive terms have $T_{\alpha'}$ and $T_{\alpha''}$ 
proportional to $T_{\alpha}$. This is equivalent to 
$T_\alpha(\rho)=t_\alpha\rho t_\alpha^*$ for some operator 
$t_\alpha:\HH_1\to\HH_2$, i.e., $T_\alpha$ is a single Kraus 
summand in a Kraus representation
 \begin{equation}\label{Krausrep} 
    T(\rho)=\sum_\alpha t_\alpha \,\rho\, t_\alpha^*\;, \qquad
    \sum_\alpha t_\alpha^*\,t_\alpha=\idty\;.
\end{equation}
Such a representation is {\it not} unique, and we will see examples later
in which the proper choice of the $t_\alpha$ is essential. However, it is easy to see that
different Kraus representations are related by \cite[Theorem 8.2]{Nielsen}
\begin{equation}\label{Krausprime}
  t_\alpha=\sum_\beta u_{\alpha\beta}\,s_\beta\;,
\end{equation}
where the two sets of Kraus operators $t_\alpha$ and $s_\beta$ have been brought to the same length by
appending zeros, and $u$ is a unitary matrix. Therefore non-refinable decompositions are given by Kraus
representations and all the others can be obtained by grouping some terms in some Kraus representation.

Finally, as announced, we prove that all realizations of a given channel with an initially pure
environment are equivalent, because every decomposition of the channel can be realized by a
measurement on the environment. Except for Section~\ref{mixedenv} we will assume from now on that
the environment is initially pure.

\begin{The}Let $U:\HH_1\otimes\KK_1\to\HH_2\otimes\KK_2$ be a unitary operator, let
$\rho_0=\ketbra{\Psi_0}{\Psi_0}\in\BB(\KK_1)$ be a pure state and let $T$ be the channel \eqref{TfromU}.
Then \emph{every} decomposition of the channel
$T=\sum_\alpha T_\alpha$ with $T_\alpha$ completely positive can be realized in the form
(\ref{TafromU}) by choosing a suitable observable $M_\alpha$ on $\KK_2$.
\end{The}

\begin{proof}
It is sufficient to prove the Theorem for a Kraus decomposition \eqref{Krausrep} because, if $T_\alpha$
are sums of Kraus terms, we simply have to find the observable for the refined Kraus decomposition and
then sum the corresponding terms.

Given a decomposition  $T(\rho)=\sum_\alpha T_\alpha(\rho) 
=\sum_\alpha t_\alpha \,\rho\, t_\alpha^*$, let us find the 
observable $M_\alpha$. Take an arbitrary complete system 
$\{\chi_\beta\}$ in $\KK_2$ (i.e., not necessarily normalized 
vectors such that 
$\sum_\beta\ketbra{\chi_\beta}{\chi_\beta}=\idty$). Then the 
corresponding observable $F_\beta=\ketbra{\chi_\beta}{\chi_\beta}$ 
inserted in \eqref{TafromU} gives the Kraus decomposition of the 
channel $T(\rho)=\sum_\beta S_\beta(\rho) =\sum_\beta s_\beta 
\,\rho\, s_\beta^*$, where $\langle\psi,s_\beta\,\varphi\rangle 
 =\langle\psi\otimes\chi_\beta,U\,\varphi\otimes\Psi_0\rangle$. 
Indeed 
 \begin{equation}\nonumber
    \langle\psi_1,S_\beta(\ketbra{\varphi_1}{\varphi_2})\,\psi_2\rangle =
    \langle U\,\varphi_2\otimes\Psi_0,\psi_2\otimes\chi_\beta\rangle
    \langle\psi_1\otimes\chi_\beta,U\,\varphi_1\otimes\Psi_0\rangle =
    \langle\psi_1,s_\beta\,\ketbra{\varphi_1}{\varphi_2}\,s_\beta^*\,\psi_2\rangle\;.
\end{equation}
Since $t_\alpha$ and $s_\beta$ are Kraus operators of a same channel $T$, there exists a unitary matrix
$u$ such that \eqref{Krausprime} holds. Then also $\mu_\alpha=\sum_\beta
\bar{u}_{\alpha\beta}\,\chi_\beta$ is a complete system in $\KK_2$ and
$M_\alpha=\ketbra{\mu_\alpha}{\mu_\alpha}$ is the required observable:
\begin{equation}\nonumber\begin{split}
    \tr\Bigl(U(\ketbra{\varphi_1}{\varphi_2}\otimes\rho_0)U^*(\ketbra{\psi_2}{\psi_1}\otimes
    M_\alpha)\Bigr) &= \sum_{\beta,\gamma} \langle
    U\,\varphi_2\otimes\Psi_0,\psi_2\otimes\bar{u}_{\alpha\beta}\,\chi_\beta\rangle
    \langle\psi_1\otimes\bar{u}_{\alpha\gamma}\,\chi_\gamma,U\,\varphi_1\otimes\Psi_0\rangle \\
    &= \sum_{\beta,\gamma} \bar{u}_{\alpha\beta}u_{\alpha\gamma}
    \langle\psi_1,s_\gamma\,\ketbra{\varphi_1}{\varphi_2}\,s_\beta^*\,\psi_2\rangle
    = \langle\psi_1,t_\alpha\,\ketbra{\varphi_1}{\varphi_2}\,t_\alpha^*\,\psi_2\rangle\;.
\end{split}\end{equation}
\end{proof}

Since we assume pure $\rho_0$ in the sequel, ``choosing a measurement $M$ on the environment'' is
synonymous with ``choosing a decomposition of the channel into completely positive summands'', and
keeping in mind the task to decide the existence of  correction scheme, we will always consider Kraus
decompositions.

\subsection{The Basic Criteria for Correction}
Before classifying channels according to the existence of a correction scheme restoring
quantum or classical information, we establish criteria to say
if a certain measurememnt on the environment, i.e.\ a certain Kraus decomposition of the channel,
allows such restoring. For both kinds of information we find simple necessary and sufficient conditions
based on the operators $t_\alpha^*t_\alpha$.

\subsubsection{Criterion for Quantum Information}\label{QBasCri}
\begin{Pro}\label{QIrest} Let $T:\BB(\HH_1)\to\BB(\HH_2)$ be a 
channel. Then, for a given Kraus decomposition 
$T(\rho)=\sum_\alpha t_\alpha\,\rho\,t_\alpha^*$, there exists a 
family of channels $R_\alpha:\BB(\HH_2)\to\BB(\HH_1)$ restoring 
quantum information if and only if 
$t_\alpha^*t_\alpha=c_\alpha\idty$ for all $\alpha$, with 
$c_\alpha\geq0$, $\sum_\alpha c_\alpha=1$. \end{Pro} \begin{proof} 
Both directions are simple. As noted in Section \ref{CorrSch}, the 
existence of the $R_\alpha$ requires that $c_\alpha = 
\tr(T_\alpha(\rho))=\tr(t_\alpha^*t_\alpha\,\rho)$ for all $\rho$, 
and hence $t_\alpha^*t_\alpha=c_\alpha\idty$. 

For the converse, suppose that the last equation holds. Then $t_\alpha=\sqrt{c_\alpha}v_\alpha$,
with $v_\alpha:\HH_1\to\HH_2$ an isometry, and we can define the restoring channels
\begin{equation}\label{QrestCh}
    R_\alpha(\rho') = v_\alpha^*\,\rho'\,v_\alpha + \rho_\alpha\,\tr\Big(\rho'(\idty-v_\alpha
    v_\alpha^*)\Big)\;,
\end{equation}
where $\rho_\alpha$ are arbitrary density operators in $\BB(\HH_1)$.
\end{proof}

The structure of the correction scheme in Proposition \ref{QIrest} is very clear when $\HH_1=\HH_2$, so
that $v_\alpha$ are unitary and the second term in \eqref{QrestCh} vanishes.
The measurement on the environment decomposes $T$ into a convex
combination of unitary channels, $T(\rho)=\sum_\alpha c_\alpha\,v_\alpha\,\rho\,v_\alpha^*$, transforming
a quantum system with one of the reversible
evolutions $\rho\mapsto v_\alpha\,\rho\,v_\alpha^*$, which occur randomly with probability
$c_\alpha$. The measurement detects the transformation occurred and $R_\alpha(\rho') =
v_\alpha^*\,\rho'\,v_\alpha$ restores the initial state $\rho$. When $\dim\HH_2>\dim\HH_1$, the operators
$v_\alpha$ are isometries, $\rho\mapsto v_\alpha\,\rho\,v_\alpha^*$ still describes a reversible
transformation and the structure of the correction scheme is similar. The channel is decomposed into a
convex combination of isometric channels transforming
a quantum system with state $\rho\in\BB(\HH_1)$ into a quantum system with one of the states
$v_\alpha\,\rho\,v_\alpha^*\in\BB(\HH_2)$, which occur randomly with probability
$c_\alpha$. Once $\alpha$ is observed, the corresponding $R_\alpha$ \eqref{QrestCh} restores $\rho$. Note
that we are interested in applying $R_\alpha$ only to states $\rho'=v_\alpha\,\rho\,v_\alpha^*$ for which
the second term in \eqref{QrestCh} vanishes, but we have to introduce such a term to get a
trace preserving $R_\alpha$.

When $\dim\HH_1=\dim\HH_2=N$, the Proposition yields an 
interesting necessary condition for corrigibility:  since each 
$t_\alpha$ is proportional to a unitary, the completely chaotic 
state $\overline{\rho}=N^{-1}\idty$ is mapped by the channel to 
$T(\overline{\rho})=N^{-1}\sum_\alpha t_\alpha 
t_\alpha^*=N^{-1}\sum_\alpha 
c_\alpha=N^{-1}\idty=\overline{\rho}$. Channels with 
$\dim\HH_1=\dim\HH_2$ and 
 \begin{equation}\label{eq:DS} 
 T(\idty)=\idty
\end{equation}
 are called {\it doubly stochastic} because this is the exact analog of a well known
property of transition probability matrices: such a matrix is called doubly stochastic if the
chaotic (equi-)distribution is invariant, which is to say that both rows and columns add up to $1$.
A famous Theorem by Birkhoff then states that every doubly stochastic matrix is a convex
combination of permutation operators, which in turn would translate to ``reversible transition
matrix'', or ``unitarily implemented channels'' in the quantum case. In other words, Birkhoff's
Theorem would suggest by analogy that double stochasticity is not only necessary but even
sufficient for the conditions of the Proposition. However, the quantum analog of Birkhoff's Theorem
fails. This was shown in \cite{Landau}, to which we refer for more background and results. Even for
channels close to the identity one can find doubly stochastic channels, which are not convex
combination of unitary ones \cite{Kummer}. It turns out that the main counterexample of
\cite{Landau} for $d=3$ is also needed in our paper (see Example~4). For $d=2$, the quantum analog
of Birkoff's Theorem holds \cite{Landau}, and we give a direct proof of this fact in
Section~\ref{sec:DS2Q}.

\subsubsection{Criterion for Classical Information}\label{CBasCri}
\begin{The}\label{CIrest} Let $T:\BB(\HH_1)\to\BB(\HH_2)$ be a channel. Then, for a given basis $B$ in
$\HH_1$ and a given Kraus
decomposition $T(\rho)=\sum_\alpha t_\alpha\,\rho\,t_\alpha^*$, there exists a
family of channels $R_\alpha:\BB(\HH_2)\to\BB(\HH_1)$ restoring classical information in basis $B$
if and only if $t_\alpha^*t_\alpha$ are all diagonal in $B$.
\end{The}
\begin{proof} If $T$ is used to send classical information encoded on $B$, then the initial
state is one of the $B_x$ and, when the measurement on the environment gives $\alpha$, the (non
normalized) output state is one of the non zero $t_\alpha\,B_x\,t_\alpha^*$. Therefore it is clear that a
restoring channel $R_\alpha$ exists if and only if $t_\alpha\,B_x\,t_\alpha^*$ are orthogonal. Indeed, if
$R_\alpha$ exists, the possible $t_\alpha\,B_x\,t_\alpha^*$ have to be orthogonal because they can be
perfectly distinguished, e.g.\ by a measurement of $B$ after the application of $R_\alpha$, while, if the
$t_\alpha\,B_x\,t_\alpha^*$ are orthogonal, a measurement can distinguish them and we could obtain
$R_\alpha$ by measuring and repreparing. In this case we would employ restoring channels of the kind
\begin{equation}\label{CrestCh}
    R_\alpha(\rho') = {\sum_x}'
    \ketbra{\varphi_x}{\psi_x^{(\alpha)}}\,\rho'\,\ketbra{\psi_x^{(\alpha)}}{\varphi_x}
    + E_{(\alpha)}\,\rho'\,E_{(\alpha)}\;,
\end{equation}
where $\varphi_x$ are the vectors in the basis $B$,
the sum is over those $x$ such that $\norm{t_\alpha\varphi_x}\neq0$, while $\psi_x^{(\alpha)} =
t_\alpha\varphi_x / \norm{t_\alpha\varphi_x}$ and $E_{(\alpha)}$ is the orthogonal projection on $\{
t_\alpha\varphi_x\;|\;x=1,\ldots,\dim\HH_1\}^\perp$.

Finally it is clear that $t_\alpha\,B_x\,t_\alpha^*$ are orthogonal, i.e.\ {}
\begin{equation}\nonumber
    t_\alpha\,\ketbra{\varphi_x}{\varphi_x}\,t_\alpha^* \;
    t_\alpha\,\ketbra{\varphi_y}{\varphi_y}\,t_\alpha^* = 0\;, \qquad \forall\; x\neq y\;,
\end{equation}
if and only if $\langle\varphi_x,t_\alpha^*t_\alpha\,\varphi_y\rangle=0$ for all $x\neq y$, i.e.\
$t_\alpha^*t_\alpha$ is diagonal in $B$.

\end{proof}

Let us remark that \eqref{CrestCh} is not the only possible choice 
of $R_\alpha$: given a Kraus decomposition of $T$, in Section 
\ref{OptRec}, eq. \eqref{OrestCh0} and \eqref{OrestCh}, we will 
find a generalization of channels \eqref{QrestCh} which could 
restore classical information as well as channels \eqref{CrestCh}, 
but which would also give a correction scheme allowing optimal 
recovery of quantum information if this were sent through the 
channel.

\subsection{Classification of Examples}\label{sec:types}
In this section we summarize the properties to be investigated in the coming Section.

\begin{Def}Consider a channel $T$, realized by coupling to an environment. We call it
\begin{itemize}
\item a {\bf `Q' channel}, if QUANTUM information can be corrected, i.e., the conditions of
         Proposition~\ref{QIrest} hold.
\item a {\bf `DS' channel}, if it is DOUBLY STOCHASTIC (cf. Eq.~\eqref{eq:DS}).
\item a {\bf `A' channel}, if for  ALL bases $B$ a correction scheme exists
          (i.e., the conditions of Theorem~\ref{CIrest} hold).
\item a {\bf `S' channel}, if for SOME basis $B$ a correction scheme exists.
\item a {\bf `N' channel}, if  NONE of the above properties is asserted.
\end{itemize}
\end{Def}

The somewhat strange category `N' is needed to make the implication `S'$\Rightarrow$`N' trivial,
and hence to get the hierarchy in the following theorem. It is understood that the implications
involving `DS' only make sense when $\dim\HH_1=\dim\HH_2$. The examples of the next section
together constitute a proof of the following Theorem:

\begin{The}\label{implications} All implications
\begin{equation}\label{inclu}
  {\rm 'DS'}\Leftarrow{\rm 'Q'}\Rightarrow{\rm 'A'}\Rightarrow{\rm 'S'}\Rightarrow{\rm 'N'}
\end{equation}
are strict in general. However, for qubit channels ($\dim\HH_1=2$), we have the implications
\begin{equation}\label{inclu2}
  {\rm 'DS'}\Leftrightarrow{\rm 'Q'}\Rightarrow{\rm 'A'}\Leftrightarrow{\rm 'S'}\Leftrightarrow{\rm 'N'}
\end{equation}
\end{The}

\section{Examples}\label{sec:examples} 

\subsection{`Q' Channels}

Let us start with two remarkable examples of channels, which are known to destroy all quantum
information and for which no ordinary quantum error correcting code works.  Nevertheless they turn
out to be channels of type `Q', so that quantum information can be completely restored with the
help of a suitable measurement on the environment.

\medskip

\noindent\textbf{Example 1.} \textit{Overall state change associated with a von Neumann
measurement.} Take $\HH_1=\HH_2$ to be $N$-dimensional, choose a basis $B$ and consider, as above,
the one dimensional projections $B_\alpha$ to the basis vectors. The complete von Neumann measurement
associated to this basis results in the channel
\begin{equation}\nonumber
    T(\rho)=\sum_{\beta=1}^N B_\beta \,\rho\, B_\beta\;.
\end{equation}
Obviously, the $B_\beta$ are diagonal only in the basis $B$ itself, so this is the unique basis in
which classical information can be sent and corrected for this particular measurement (Kraus
decomposition of $T$). However, we can also represent $T$ as
\begin{equation}\nonumber
    T(\rho)=\sum_{\alpha=1}^N t_\alpha \,\rho\, t_\alpha\;, \qquad
    t_\alpha = \sum_{\beta=1}^N \frac{1}{\sqrt{N}} \e^{\frac{2\pi\im}{N}\alpha\beta} B_\beta =
    \frac{1}{\sqrt{N}} V_\alpha\;,
\end{equation}
where all the $V_\alpha$ are unitary. Hence $T$ is a `Q' channel.

This channel has quantum channel capacity zero \cite{QECC}, but classical capacity $\log_2(N)$. The
following channel is even worse: without help from the environment it allows no information
transmission at all.
\medskip

\noindent\textbf{Example 2.} \textit{Depolarising channel.}

Take $\HH_1=\HH_2$ with $\dim \mathcal{H}_1 = N$. Then the   depolarising channel is defined by
$T(\rho)= \frac{1}{N} \idty$. It is a `Q' channel because it admits the Kraus representation
\begin{equation}\nonumber
    T(\rho) = \sum_{j,k=1}^N t_{j,k} \,\rho\, t_{j,k}^*\;, \qquad
    t_{j,k} = \frac{1}{N} \sum_{x=1}^N \e^{\frac{2\pi\im}{n}xk} \ketbra{x+j}{x}
            = \frac{1}{N}\,  v_{j,k}\;,
\end{equation}
with unitary operators $v_{j,k}$, where  $\{|x\rangle\}_{x\in\mathbb{Z}_N}$ denotes a  basis
labeled cyclically so that addition in $|x+j\rangle$ is modulo $N$. The representations of the
depolarizing channel with Kraus operators proportional to unitaries,
of which there are very many for larger $N$, are
in fact, in one-to-one correspondence to quantum teleportation schemes and superdense coding
schemes \cite{telepo}.

\medskip

\textbf{Casimir channel.} The next example and some of the following belong to a family we call
Casimir channels. These can be defined starting from any compact Lie group. Let us recall that all
irreducible unitary representations $\pi$ of such a group live in finite  dimensional Hilbert spaces
$\HH_\pi$ and that its Lie algebra admits an invariant positive definite quadratic form $g$
\cite{BarRac}. Then it is easy to define a channel with input and output spaces equal to $\HH_\pi$:
if $L_\alpha$ denote the generators of the Lie algebra and if we consider the invariant quadratic
polynomial $\sum_{\alpha,\beta} g_{\alpha\beta}\,L_\alpha\,L_\beta$ (second order Casimir
operator), then $\sum_{\alpha,\beta} g_{\alpha\beta} \,\pi(L_\alpha) \,\pi(L_\beta) =
\lambda_\pi\idty$, because this operator commutes with the group representation, and $\pi$ is
irreducible. We use here the convention that $\pi(L)$ denotes the self-adjoint (rather than
skew-adjoint) generators of the group in the representation $\pi$. Then $\lambda_\pi>0$, and the
map
\begin{equation}\nonumber
    T(\rho) = \sum_{\alpha,\beta} \frac{1}{\lambda_\pi} g_{\alpha\beta}
    \,\pi(L_\alpha) \,\rho\, \pi(L_\beta)
\end{equation}
defines automatically a channel, which is completely positive because we could choose generators
$L_\alpha$ diagonalizing $g$ and hence giving directly a Kraus representation of $T$. Note that
this is always a `DS' channel because the latter Kraus operators are self-adjoint.

In examples 3, 4 and 6 we consider Casimir channels associated with irreducible representations of
$\SU2$, with $g_{\alpha\beta} = \delta_{\alpha\beta}$ and $J_\alpha=\pi(L_\alpha)$ equal to the
angular momentum operators generating a spin-$s$ representation of $\SU2$ on $\HH$, $\dim\HH=2s+1$.
Then $\sum_\alpha J_\alpha^2=s(s+1)\idty$ and the channel is
\begin{equation}\label{CasimirCh}
    T(\rho)=\sum_{\alpha=1}^3 \frac{1}{s(s+1)} J_\alpha \,\rho\, J_\alpha\;.
\end{equation}

\medskip

\noindent\textbf{Example 3.} \textit{Casimir channel}, $s=1/2$. Take $\HH_1=\HH_2=\Cx^2$, and use
the Pauli spin operators $\pi(L_\alpha)=\frac12\sigma_\alpha$ as generators of the spin-$\frac12$
representation of $\SU2$. This gives the Casimir channel ($\lambda_\pi=\frac12(\frac12+1)=\frac34$)
\begin{equation}\nonumber
    T(\rho)=\sum_{\alpha=1}^3 \frac{1}{3} \sigma_\alpha \,\rho\, \sigma_\alpha =
    \frac{2}{3}\idty - \frac{1}{3}\rho\;,
\end{equation}
which is a `Q' channel because the Pauli matrices $\sigma_\alpha$ are unitary. This is the best
completely positive approximation to the ``Universal Not'' operation, which would take any pure
state on $\Cx^2$ to its orthogonal complement \cite{unot}.

\subsection{`A not Q' Channels}
We give two examples of `A not Q' channels. Example 4 is `DS', and hence also a counterexample to
the quantum analog of Birkhoff's Theorem. Example 5 is not `DS', so property `Q' may fail in
different ways.

\medskip

\noindent\textbf{Example 4.} \textit{Casimir channel}, $s=1$. This representation is just the
orthogonal group in 3 dimensions. Hence we take $\HH_1=\HH_2=\Cx^3$ and, with a basis
$\{|x\rangle\}_{x=1}^3$, we can write
\begin{equation}\nonumber
    T(\rho)=\sum_{\beta=1}^3 \frac{1}{2} J_\beta \,\rho\, J_\beta\;,
    \qquad \langle i|J_\beta|j\rangle = \im \varepsilon_{ij\beta}\;,
\end{equation}
 where $\varepsilon_{ij\beta}$ is the completely antisymmetric tensor and $J_\beta$ 
are the angular momentum operators in spin-1 representation. This 
channel was already used as a counterexample to the quantum analog 
of  Birkhoff's Theorem in \cite{Landau}, but also to a completely 
different problem in \cite{holevoW}.

Since the $J_\beta$ are antisymmetric, the same will be true for 
any choice of Kraus operators, which must be of the form 
$t_\alpha=\sum_\beta u_{\alpha\beta}\,J_\beta/\sqrt{2}$ by 
eq.~(\ref{Krausprime}). But in odd dimension an antisymmetric 
matrix necessarily has determinant zero, and hence cannot be 
unitary. It follows that $T$ is not a `Q' channel. On the other 
hand, like every Casimir channel, $T$ enjoys  property `DS'. 

It remains to show that $T$ is an `A' channel. Note that the $J_\alpha^2$ are diagonal, so it is
obvious that correction works for the standard basis and, by $\SU2$ invariance, for all bases
arising from this by orthogonal (\emph{real} unitary) transformation.  But consider an arbitrary
orthonormal basis  $\{\varphi_y\}_{y=1}^3$. Then we can take the unitary matrix rotating the
standard basis to $\{\varphi_y\}$  also as the matrix transforming the Kraus operators. Then
\begin{equation}\nonumber
    T(\rho)=\sum_\alpha t_\alpha \,\rho\, t_\alpha\;, \qquad
    t_\alpha = \sum_\beta \langle\varphi_\alpha|\beta\rangle \, \frac{1}{\sqrt{2}}J_\beta\;.
\end{equation}
Using the properties $J_\beta^2=\idty-\ketbra{\beta}{\beta}$ and, for $\beta\neq\gamma$, $J_\beta
J_\gamma=-\ketbra{\gamma}{\beta}$, it is easy to verify that
\begin{equation}\nonumber
    t_\alpha^*t_\alpha = \frac{1}{2} \Big( \idty - \ketbra{\varphi_\alpha}{\varphi_\alpha} \Big)\;.
\end{equation}
Since these operators are diagonal w.r.t.\ $\{\varphi_y\}$, $T$ is an `A' channel.

\medskip

\noindent\textbf{Example 5.}\textit{Collapsing channel}

Let $\dim\HH_1>1$, fix an arbitrary unit vector $\psi$ in $\HH_2$ and define the channel
\begin{equation}\nonumber
    T(\rho) = \ketbra{\psi}{\psi}\;.
\end{equation}
Then $T$ is not a `Q' channel because every Kraus operator $t_\alpha$ necessarily has rank 1, and of
course, if $\dim\HH_1=\dim\HH_2$, it is neither a `DS' channel. Anyway it is always an `A' channel.
Indeed for every basis $\{\varphi_x\}$ in $\HH_1$ we have the Kraus representation
\begin{equation}\nonumber
    T(\rho) = \sum_\alpha t_\alpha \,\rho\, t_\alpha^* =
    \sum_\alpha \ketbra{\psi}{\varphi_\alpha} \,\rho\, \ketbra{\varphi_\alpha}{\psi}\;,
\end{equation}
where all $t_\alpha^*t_\alpha = \ketbra{\varphi_\alpha}{\varphi_\alpha}$ are diagonal w.r.t.\
$\{\varphi_x\}$.

\subsection{`S not A' Channels}

\noindent\textbf{Example 6.} \textit{Casimir channel}, $s=3/2$. We show first that {\it every}
$\SU2$-Casimir channel is an `S' channel. Indeed, we can rewrite the channel \eqref{CasimirCh} in
the Kraus representation
\begin{equation}\nonumber
    T(\rho) = \sum_{\alpha=1}^3 t_\alpha \,\rho\, t_\alpha^*, \qquad
    t_1=\frac{J_1+\im J_2}{\sqrt{2s(s+1)}} = \frac{J_+}{\sqrt{2s(s+1)}}, \quad
    t_2=\frac{J_1-\im J_2}{\sqrt{2s(s+1)}} = \frac{J_-}{\sqrt{2s(s+1)}}, \quad
    t_3=\frac{J_3}{\sqrt{s(s+1)}},
\end{equation}
where
\begin{equation}\nonumber
    t_1^*t_1 = \frac{1}{2s(s+1)} J_-J_+, \quad
    t_2^*t_2 = \frac{1}{2s(s+1)} J_+J_-, \quad
    t_3^*t_3 = \frac{1}{s(s+1)} J_3^2
\end{equation}
are all diagonal w.r.t.\ the eigenbasis of $J_3$.

On the other hand, the channel with $s=3/2$ is not of type `A', because there is no non-zero linear
combination $t = xJ_1+yJ_2+zJ_3$ such that $t^*t$ is diagonal w.r.t.\ the basis
\begin{eqnarray}
 \varphi_1&=&(|3/2\rangle+\im|1/2\rangle)/\sqrt{2}\;,   \nonumber\\
 \varphi_2&=&(|3/2\rangle-\im|1/2\rangle)/\sqrt{2}\;,   \nonumber\\
 \varphi_3&=&|-1/2\rangle\;,   \nonumber\\
 \varphi_4&=&|-3/2\rangle\;,   \nonumber
\end{eqnarray} where $|m\rangle$ denotes the eigenstates of $J_3$.

Let us note that examples 5 and 6 together show that there is no logical relation between
properties `A' and `DS'.

\medskip

\subsection{`N not S' Channels}

\noindent\textbf{Example 7.} In order to exhibit a channel $T$ without even property `S', we
construct Kraus operators $t_\alpha:\HH\to\HH$, $\alpha=1,\ldots,3$, such that
\begin{equation}\label{comm0}
    \Bigl[|\sum_\alpha\xi_\alpha t_\alpha|^2,|\sum_\alpha\zeta_\alpha t_\alpha|^2\Bigr]=0
\end{equation}
implies linear dependence of the vectors $\xi$ and $\zeta$. Then no matter how we recombine the
Kraus operators $t_\alpha$ according to eq.~(\ref{Krausprime}), we cannot make even two (let alone all)
$s_\alpha^*s_\alpha$ commute, so they cannot be jointly diagonal in any basis.

The idea of the construction is to take the Kraus operators as direct sums
$t_\alpha=\bigoplus_{i=0}^N t_\alpha^{(i)}$ such that for every $i$, the $t_\alpha^{(i)}$ are Kraus
operators in their own right. Then eq.~(\ref{comm0}) becomes
\begin{equation}\nonumber
  \bigoplus_{i=0}^N \;
     \Bigl[|\sum_\alpha\xi_\alpha t_\alpha^{(i)}|^2,|\sum_\alpha\zeta_\alpha t_\alpha^{(i)}|^2\Bigr]
     = 0\;.
\end{equation}
This holds if and only if (\ref{comm0}) holds for every summand. Hence we can increase the demands
on $\xi$ and $\zeta$ by adding more and more terms to the direct sum, up to leaving only the obvious
solution $\xi\propto\zeta$.

It is convenient to choose one summand as $t_\alpha^{(0)} = \ketbra{\psi}{\alpha}$, where
$\psi\in\Cx^3$ is an arbitrary unit vector, and the $|\alpha\rangle$ are a basis in $\Cx^3$. For
this choice $\sum_\alpha \xi_\alpha t_\alpha=\ketbra{\psi}{\overline{\xi}}$, where $\overline{\xi}$
denotes the vector with the complex conjugate components. Then the commutator (\ref{comm0}) becomes
$\bigl[\ketbra{\overline{\xi}}{\overline{\xi}},\ketbra{\overline{\zeta}}{\overline{\zeta}}\bigr]$,
which vanishes iff $\xi$ and $\zeta$ are either linearly dependent or orthogonal.

We therefore consider the following set of vector pairs in $\mathbb{C}^3\times\mathbb{C}^3$:
\begin{equation}\nonumber
    \mathcal{M} = \Bigl\{ (\xi,\zeta)\in\mathbb{C}^3\times\mathbb{C}^3 \bigm|
    \langle\xi,\zeta\rangle=0, \; \norm{\xi}=\norm{\zeta}=1 \Bigr\}\;.
\end{equation}
Our problem is reduced to showing that there is a finite collection of Kraus operators
$\{t_\alpha^{(i)}\}$, $i=1,\ldots,N$ such that the commutators (\ref{comm0}) do not vanish
simultaneously at the same point in $\cal M$.

It is easy to see that for any pair $(\xi,\zeta)\in\cal M$ we can find Kraus operators $\tilde{t}_\alpha$
for which the commutator is non-zero. For example we can pick a $3\times3$ unitary matrix
$u$ whose first two rows are $\xi$ and $\zeta$, and set
$\tilde t_\alpha = \sum_\beta \bar{u}_{\beta\alpha}\frac{2}{\sqrt{15}}J_\beta$, where $J_\beta$ are the
angular momentum operators in spin-3/2 representation, so getting $\sum_\alpha\xi_\alpha
\tilde{t}_\alpha=J_1$ and $\sum_\alpha\zeta_\alpha \tilde{t}_\alpha=J_2$. Once we have found
$\tilde{t}_\alpha$ giving non zero commutator for a certain pair $(\xi,\zeta)$, the same operators will
work for all $(\xi',\zeta')$ in a small open
neighbourhood of $(\xi,\zeta)$. But from this open covering of the compact set $\cal M$ we can
select a finite subcover, and the corresponding Kraus operators form the desired direct sum
channel.

Of course, this construction is non-constructive in the sense that we get no bound on the dimension
of the Hilbert space $\HH$. A more
explicit example, perhaps in 3 dimensions with 3 Kraus operators would clearly be preferable.

\section{Qubit Channels}\label{sec:qubits} 

As stated in Theorem \ref{implications}, qubit channels are special from the point of view of our
classification because they enjoy at least property `A' and because double stochasticity becomes
equivalent to property `Q'. We are now going to prove these results.

\subsection{`N' $\Rightarrow$ `A'}

Here we have to establish property `A' for all qubit channels $T:\BB(\HH_1)\to\BB(\HH_2)$ with
$\HH_1=\Cx^2$ the two dimensional Hilbert space. So consider an arbitrary basis of
$\HH_1$, which we can take to be  $\{|0\rangle,|1\rangle\}$. We have to show that there exist a Kraus
representation $T(\rho) = \sum_\alpha t_\alpha \,\rho\, t_\alpha^*$ of the channel $T$ such that
$t_\alpha^*t_\alpha$ are all diagonal, i.e.\ such that $\langle0|t_\alpha^*t_\alpha|1\rangle=0$ for
all $\alpha$.

We assume that the arbitrary channel $T$ has Kraus representation $T(\rho) = \sum_\beta s_\beta
\,\rho\, s_\beta^*$. Clearly then $\langle0|s_\beta^*s_\beta|1\rangle$ does not need to vanish, but
$\sum_\beta \langle0|s_\beta^*s_\beta|1\rangle = \langle0|\idty|1\rangle=0$. Therefore, if we
introduce the matrix $X_{\alpha\beta} = \langle0|s_\alpha^*s_\beta|1\rangle$, it has $\tr(X)=0$.
Rotating the Kraus operators by a unitary matrix $u$ (cf. (\ref{Krausprime})) is the same as
transforming $X\mapsto \bar{u}X\bar{u}^*$. What we want to achieve is to make the diagonal of
$\bar{u}X\bar{u}^*$ vanish
identically. That we can always find such a unitary transformation is the content of the following
Lemma.

\begin{Lemma}
Let $X$ be an operator on a finite dimensional Hilbert space $\HH$ with $\tr X=0$. Then there is an
orthonormal basis $\{e_\alpha\}$ such that $\langle e_\alpha,X\,e_\alpha \rangle=0$ for all
$\alpha$.
\end{Lemma}

This lemma already appeared in \cite{WSHV} for $\dim\HH=2^k$, but we need it for arbitrary
$\dim\HH=N$, and our proof is rather different.

\medskip

\begin{proof}
It is sufficient to show that there exists a vector $\varphi\neq0$ such that
$\langle\varphi,X\,\varphi\rangle=0$. Indeed, we can set $e_1=\varphi/\norm{\varphi}$, consider
$\varphi^\perp$, the subspace orthogonal to $\varphi$, and then repeat the same argument for
$X_1=P_{\varphi^\perp} X|_{\varphi^\perp}$, the operator on $\varphi^\perp$ obtained by restricting
$X$ to $\varphi^\perp$ and then projecting the image on $\varphi^\perp$: $X_1$ is still a null
trace operator and by induction we can find $e_2$ and the rest of the basis.

Let us show the existence of $\varphi$. Set $A=(X+X^*)/2$ and $B=(X-X^*)/2\im$ so that $X=A+\im B$,
with $A$ and $B$ null trace self-adjoint operators, and consider the eigenbasis $\{|k\rangle\}$ of
$A=\sum_k a_k \ketbra{k}{k}$. Then for every $\varphi\in\HH$, $\varphi=\sum_k\varphi_k|k\rangle$,
\begin{equation}\nonumber
    \langle\varphi,X\,\varphi\rangle = \sum_k a_k |\varphi_k|^2
    + \im\langle\varphi,B\,\varphi\rangle\;,
\end{equation}
where the first summand vanishes for $|\varphi_k|$ constant. Therefore, if we consider in $\HH$ the
torus of unit vectors
\begin{equation}\nonumber
    \mathcal{T} = \Big\{ \varphi\in\HH \Big| \varphi_k=\frac{\e^{\im \theta_k}}{\sqrt{N}},
    \quad \theta_k\in [0,2\pi), \quad k=1,\ldots,N \Big\}\;,
\end{equation}
we have that
\begin{equation}\nonumber
    \phi\mapsto \langle\varphi,X\,\varphi\rangle = \im\langle\varphi,B\,\varphi\rangle
    = \frac{\im}{N}\sum_{\ell,k} \e^{-\im(\theta_\ell-\theta_k)}\langle\ell|B|k\rangle\;,
\end{equation}
is a purely immaginary valued continuous function on $\cal T$, with average
\begin{equation}\nonumber
    \frac{1}{N}\sum_{\ell,k}\ (2\pi)^{-N} \int_0^{2\pi} \dd\theta_1 \cdots \int_0^{2\pi} \dd\theta_N
    \e^{-\im(\theta_\ell-\theta_k)}\langle\ell|B|k\rangle = \frac{\tr B}{N} = 0\;.
\end{equation}
Hence there is at least one unit vector $\varphi\in\mathcal{T}$ such that
$\langle\varphi,X\,\varphi\rangle = \im\langle\varphi,B\,\varphi\rangle=0$.

\end{proof}

\subsection{`DS' $\Rightarrow$ `Q'}\label{sec:DS2Q}

It was already shown in \cite{Landau} that for qubits the quantum 
analog of Birkhoff's Theorem holds. The proof given in that paper 
was based, however, on a full analysis of the extremal doubly 
stochastic operators. Here we give a different, more direct proof. 
 
Given a doubly stochastic qubit channel $T:\BB(\HH)\to\BB(\HH)$, 
with $\HH=\Cx^2$ the two dimensional Hilbert space, we have to 
show that there exist a Kraus representation $T(\rho) = 
\sum_\alpha t_\alpha \,\rho\, t_\alpha^*$ such that $t_\alpha$ are 
all proportional to unitary operators. Let us recall that 
$u\in\BB(\HH)$, $\dim\HH=2$, is unitary if and only if 
$u=\e^{\im\theta}(a_0\idty+\im\sum_{j=1}^3a_j\sigma_j)$ for real 
$\theta$, $a_0,\ldots,a_3$, such that $\sum_ja_j^2=1$. As usual, 
$\sigma_j$ denote the Pauli matrices with $\sigma_0=\idty$. By 
expanding all Kraus operators in Pauli matrices we can write the 
given doubly stochastic channel as 

\begin{equation}\nonumber
    T(\rho) = \sum_{i,j=0}^3 R_{ij} \, \sigma_i \,\rho\, \sigma_j\;,
\end{equation}
with some complex $4\times4$-matrix $R=(R_{ij})$ enjoying the properties
\begin{equation}\label{Rprop}
    R\geq0, \qquad \tr R=1, \qquad R_{0j}=-R_{j0}, \quad R_{ij}=R_{ji}, \quad \forall\; i,j=1,2,3.
\end{equation}
Property $R\geq0$ is equivalent to the complete positivity of $T$, while the others are equivalent
to $\tr T(\rho)=\tr\rho$ and $T(\idty)=\idty$. We need the following lemma.

\begin{Lemma}\label{DSiQ}
Let $R=(R_{ij})_{i,j=0}^3$ be a $4\times4$ complex matrix enjoying properties \eqref{Rprop}. Then
there exists an orthonormal basis of eigenvectors $\varphi^{(\alpha)}\in\Cx^4$ of $R$ with the
first component $\varphi^{(\alpha)}_0\in\mathbb{R}$ and the other
$\varphi^{(\alpha)}_j\in\im\mathbb{R}$.
\end{Lemma}

Before proving this lemma, let us verify that it does solve our 
problem. Indeed, if the $\varphi^{(\alpha)}$ are such a basis, use 
the spectral decomposition  $R=\sum_\alpha r_\alpha 
\ketbra{\varphi^{(\alpha)}}{\varphi^{(\alpha)}}$, $r_\alpha\geq0$,  
to get 
\begin{equation} 
    T(\rho) = \sum_{i,j=0}^3 R_{ij} \, \sigma_i \,\rho\, \sigma_j
        =  \sum_{i,j=0}^3 \sum_{\alpha=0}^3 r_\alpha \varphi^{(\alpha)}_i
                 \bar\varphi^{(\alpha)}_j \, \sigma_i \,\rho\, \sigma_j 
            = \sum_{\alpha=0}^3 r_\alpha \Big(\sum_{i=0}^3  \varphi^{(\alpha)}_i
                    \, \sigma_i \Big) \rho \Big( \sum_{j=0}^3 
                         \bar\varphi^{(\alpha)}_j \, \sigma_j \Big)
            = \sum_{\alpha=0}^3 t_\alpha \,\rho\, t_\alpha^*\;,  \nonumber
\end{equation}
where $t_\alpha= \sqrt{r_\alpha} \Big(\sum_{i=0}^3 \varphi^{(\alpha)}_i \, \sigma_i \Big)$
with $\sum_{i=0}^3 \varphi^{(\alpha)}_i \, \sigma_i$ unitary.

\medskip

\begin{proof}
If we introduce the antiunitary involution $S:\mathbb{C}^4\to\mathbb{C}^4$
\begin{equation}\nonumber
    S\begin{pmatrix} \varphi_0 \\ \varphi_1 \\ \varphi_2 \\ \varphi_3 \end{pmatrix}
    =\begin{pmatrix}\bar\varphi_0 \\ -\bar\varphi_1 \\ -\bar\varphi_2 \\ -\bar\varphi_3\end{pmatrix},
\end{equation}
then the conditions \eqref{Rprop} on $R$
become $R\geq0$ and $S\circ R=R\circ S$, whereas the required property for the eigenvectors $\varphi$
can be written as $S\varphi=\varphi$. Therefore we are given $S\circ R=R\circ S$ and we are looking for
``real'', i.e.\ $S$-invariant, eigenvectors.

Consider any eigenvector $\varphi$ of $R$. Then $S\varphi$ is an 
eigenvector for the same eigenvalue. Now we can either have that 
$S\varphi$ is proportional to $\varphi$, in which case we can 
adjust the phase to make $S\varphi=\varphi$. Or else, $S\varphi$ 
and $\varphi$ are linearly independent, and span a two-dimensional 
subspace which is invariant under $S$, and on which $R$ acts like 
a positive multiple of $\idty$. In this subspace we can again find 
a basis of $S$-invariant vectors. In either case, the orthogonal 
complement of the subspace generated by $\varphi$ and $S\varphi$ 
is again invariant under both operators, and we continue with the 
eigenvectors of $R$ in this complement, thereby constructing 
successively a basis of $S$-invariant eigenvectors. 

\end{proof}

\section{Optimal recovery of Quantum Information}\label{OptRec}

 If quantum information is sent through a `not Q' channel, it is  
natural in our framework to look for the correction schemes which 
give the best possible information preserving, i.e.\ which bring 
the corrected channel $\Tcorr$  \eqref{Tcorr} as close as possible 
to $\id_1$, in some sense. In this Section we find optimal 
restoring channels $R_\alpha$ for a given measurement $M_\alpha$. 
Even if the result leaves unsolved the problem of choosing 
$M_\alpha$, it is interesting by itself for all those situations 
where practical constraints could prevent the choice of $M_\alpha$ 
(which could happen also with a `Q' channel). We define the 
optimal $R_\alpha$ to be the channels which maximize the 
\emph{channel fidelity} of the corrected channel, $\fid(\Tcorr)$. 
For a channel $T:\BB(\HH)\to\BB(\HH)$, $T(\rho)=\sum_\alpha 
t_\alpha\,\rho\,t_\alpha^*$, denoted by $\Omega$ a maximally 
entangled unit vector in $\HH\otimes\HH$, the channel fidelity 
\begin{equation}\nonumber 
    \fid(T) = \langle\Omega,T\otimes\id(\ketbra{\Omega}{\Omega})\,\Omega\rangle
    = \frac{1}{(\dim\HH)^2} \sum_\alpha |\tr t_\alpha|^2
\end{equation}
measures how well $T$ preserves quantum information, reaching 1 if and only if $T=\id$.

Given $T:\BB(\HH_1)\to\BB(\HH_2)$, chosen a Kraus decomposition \eqref{Krausrep} and the restoring
channels $R_\alpha:\BB(\HH_2)\to\BB(\HH_1)$,
\begin{equation}\nonumber
    R_\alpha(\rho')=\sum_\beta r_\beta^\alpha \,\rho\, {r_\beta^\alpha}^*\;, \qquad
    \sum_\beta {r_\beta^\alpha}^*\,r_\beta^\alpha=\idty\;,
\end{equation}
we are interested in
\begin{equation}\label{FidVal}
    \fid(\Tcorr) = \frac{1}{(\dim\HH_1)^2} \sum_{\alpha,\beta} |\tr r_\beta^\alpha t_\alpha|^2\;,
\end{equation}
and we want to maximize it with respect to all possible families $\{R_\alpha\}$.
\begin{Pro} Let $T:\BB(\HH_1)\to\BB(\HH_2)$ be a channel with a fixed Kraus decomposition
\eqref{Krausrep} and, for every family of channels $R_\alpha:\BB(\HH_2)\to\BB(\HH_1)$, let
$\Tcorr:\BB(\HH_1)\to\BB(\HH_1)$ be the corresponding overall corrected channel \eqref{Tcorr}. Then
for every family $\{R_\alpha\}$
\begin{equation}\label{FidMaj}
    \fid(\Tcorr) \leq \frac{1}{(\dim\HH_1)^2} \sum_\alpha \Big(\tr |t_\alpha| \Big)^2\;.
\end{equation}
Moreover equality  can always be attained and it holds if and only if
\begin{equation}\label{OrestCh0}
    R_\alpha(\rho') = v_\alpha^*\,\rho'\,v_\alpha + \rin(\rho'), \qquad \mbox{ where }
    \rin(t_\alpha\,\rho\,t_\alpha^*) = 0 \quad \forall\; \rho\in\BB(\HH_1),
\end{equation}
or, equivalently, if and only if
\begin{equation}\label{OTcorr}
    \Tcorr(\rho) = \sum_\alpha |t_\alpha| \,\rho\, |t_\alpha| \;.
\end{equation}
\end{Pro}

Before proving the Proposition we write an explicit example of channels \eqref{OrestCh0}. To do this,
we need to introduce the polar decomposition of the Kraus operators of $T$,
\begin{equation}\label{PolarDec}
    t_\alpha = v_\alpha \, |t_\alpha|\;,
\end{equation}
where $|t_\alpha| = \sqrt{t_\alpha^*t_\alpha}:\BB(\HH_1)\to\BB(\HH_1)$ and $v_\alpha$ is a partial
isometry from $\range(|t_\alpha|)$ to $\range(t_\alpha)$ which we extend to an operator from
$\HH_1$ to $\HH_2$ by defining it to be 0 on $\range(|t_\alpha|)^\perp$. Then $v_\alpha^*v_\alpha$ is
the orthogonal projection on $\range(|t_\alpha|)$ while $v_\alpha v_\alpha^*$ is
the orthogonal projection on $\range(t_\alpha)$. Using $v_\alpha$ we can satisfy \eqref{OrestCh0}
defining the channels
\begin{equation}\label{OrestCh}
    R_\alpha(\rho') = v_\alpha^*\,\rho'\,v_\alpha + \rho_\alpha\,\tr\Big(\rho'(\idty-v_\alpha
    v_\alpha^*)\Big)\;,
\end{equation}
where $\rho_\alpha$ are arbitrary density operators in $\BB(\HH_1)$.

\medskip

\begin{proof} Inequality \eqref{FidMaj} follows from \eqref{FidVal} because the
Cauchy-Schwarz inequality for the Hilbert-Schmidt inner product gives
\begin{equation}\label{CS}
    \sum_\beta |\tr r_\beta^\alpha t_\alpha|^2
    = \sum_\beta\Big|\tr \big(r_\beta^\alpha v_\alpha|t_\alpha|^{1/2} |t_\alpha|^{1/2}\big) \Big|^2
    \leq \sum_\beta \tr \big(|t_\alpha|^{1/2}v_\alpha^*{r_\beta^\alpha}^*r_\beta^\alpha
    v_\alpha|t_\alpha|^{1/2}\big) \, \tr|t_\alpha|
    = \Big( \tr|t_\alpha|\Big)^2 \;.
\end{equation}
If $R_\alpha$ are chosen according to \eqref{OrestCh}, then $\Tcorr$ is given by \eqref{OTcorr} and
equality holds in \eqref{FidMaj}. And the same is true for every family of restoring channels
\eqref{OrestCh0}.

Still remaining to prove is that equality in \eqref{FidMaj} implies \eqref{OrestCh0}.
Equality holds in \eqref{FidMaj} if and only if equality holds in \eqref{CS} for every $\alpha$,
which occurs if and only if $r_\beta^\alpha v_\alpha|t_\alpha|^{1/2} =
\lambda_{\alpha\beta}|t_\alpha|^{1/2}$, $\lambda_{\alpha\beta}\in\mathbb{C}$, for
every $\alpha$ and $\beta$. Then $r_\beta^\alpha t_\alpha = \lambda_{\alpha\beta}t_\alpha$ and
$\rin(t_\alpha\,\rho\,t_\alpha^*) = R_\alpha(t_\alpha\,\rho\,t_\alpha^*) - v_\alpha^* \,
t_\alpha\,\rho\,t_\alpha^* \, v_\alpha = 0$.

\end{proof}

The structure of the optimal restoring channels \eqref{OrestCh0} or \eqref{OrestCh}, obtained by
maximizing $\fid(\Tcorr)$, is just what one could expect. When a measurement on the environment has given
the result $\alpha$, we deal with a subensemble of systems undergone the state transformation
$\rho\mapsto t_\alpha\,\rho\,t_\alpha^* = v_\alpha |t_\alpha|\,\rho\,|t_\alpha|v_\alpha^*$, which can be
seen as a composition of $\rho\mapsto |t_\alpha|\,\rho\,|t_\alpha|$ followed by $\rho\mapsto
v_\alpha\,\rho\,v_\alpha^*$. Unless we are in the trivial case $|t_\alpha|\propto\idty$, only the second
transformation is physically reversible, and this is just what the channels \eqref{OrestCh0} and
\eqref{OrestCh} do.
Notice that if $|t_\alpha|\propto\idty$ then $T$ is decomposed into isometric channels, the restoring
channels \eqref{OrestCh} coincide with \eqref{QrestCh}, the correction scheme restores quantum
information and $\fid(\Tcorr)=1$.
As anticipated in Section \ref{CBasCri}, a correction scheme based on channels \eqref{OrestCh0} or
\eqref{OrestCh} not only recovers optimally quantum information, but if the measurement on the
environment allows restoring of classical information in a basis $B$, i.e.\ if $|t_\alpha|$ are diagonal
in $B$, then it also restores such information.

\section{Mixed Environments}\label{mixedenv}

When a channel \eqref{TfromU} is realised by coupling the system 
to an environment whose initial state $\rho_0$ is not pure, the 
possible decompositions of the channel $T=\sum_\alpha T_\alpha$ 
given by a measurement on the environment (see \eqref{TafromU}) 
actually depend on $\rho_0$. As a consequence, the possibility of 
restoring quantum or classical information sent through the 
channel does not depend only on the channel, but also on the 
details of the coupling. We will illustrate this feature with an 
example of a `N not S' depolarizing qubit channel. Since the same 
channel would be `Q' if the environment were pure, the example 
shows that a mixed environment can cause the highest possible drop 
in our hierarchy. Moreover, since it is a qubit channel, it also 
shows that with mixed environments the special properties of 
qubits are destroyed.  

\medskip \noindent\textbf{Example 8.} Let us consider the channel 
$T:\BB(\HH)\to\BB(\HH)$ arising from coupling a qubit, of Hilbert 
space $\HH$, to another qubit, the environment, of Hilbert space 
$\KK$ and initial state $\rho_0=\idty/2$, by the interaction $U$ 
\begin{equation}\nonumber 
    U\,\chi_1\otimes\xi_1=\psi_1\otimes\eta_1, \qquad
    U\,\chi_1\otimes\xi_0=\psi_0\otimes\zeta_1, \qquad
    U\,\chi_0\otimes\xi_1=\psi_1\otimes\eta_0, \qquad
    U\,\chi_0\otimes\xi_0=\psi_0\otimes\zeta_0,
\end{equation} 
 where $\{\chi_x\}\subset\HH$, $\{\psi_x\}\subset\HH$, $\{\xi_x\}\subset\KK$, and 
$\{\eta_x\}\subset\KK$ are arbitrary orthonormal bases. The 
coupling is non-trivial, because it contains yet another basis 
$\{\zeta_x\}\subset\KK$, explicitly given by 
 \begin{equation}\nonumber 
    \zeta_1=\frac{\eta_1+\eta_0}{\sqrt2}, \qquad \zeta_0=\frac{\eta_1-\eta_0}{\sqrt2}.
\end{equation}
 Then it is easy to verify that \eqref{TfromU}  
gives the depolarizing channel $T(\rho)=\idty/2$. But, even if the 
evolution $U$ is followed by the measurement on the second qubit 
of an observable $M_\alpha=\ketbra{\mu_\alpha}{\mu_\alpha}$, for 
some complete system $\{\mu_\alpha\}$ in $\KK$, anyhow the 
classical information sent through this channel is lost, for every 
choice of the encoding basis $B$: the (non normalized) states 
$T_\alpha(B_1)$ and $T_\alpha(B_0)$ are always non orthogonal and 
hence no channel $R_\alpha$ can distinguish them and restore the 
initial states. Indeed, for every basis $B=\{\varphi_x\}$ in 
$\HH$, 
\begin{equation}\nonumber\begin{array}{ll} 
    U\,\varphi_1\otimes\xi_1 =
    \psi_1\otimes\Big(\langle\chi_1,\varphi_1\rangle\eta_1+\langle\chi_0,\varphi_1\rangle\eta_0\Big)
    = \psi_1\otimes\tilde{\eta}_1, \qquad &
    U\,\varphi_1\otimes\xi_0 =
    \psi_0\otimes\Big(\langle\chi_1,\varphi_1\rangle\zeta_1+\langle\chi_0,\varphi_1\rangle\zeta_0\Big)
    =\psi_0\otimes\tilde{\zeta}_1, \\
    U\,\varphi_0\otimes\xi_1 =
    \psi_1\otimes\Big(\langle\chi_1,\varphi_0\rangle\eta_1+\langle\chi_0,\varphi_0\rangle\eta_0\Big)
    =\psi_1\otimes\tilde{\eta}_0, \qquad &
    U\,\varphi_0\otimes\xi_0 =
    \psi_0\otimes\Big(\langle\chi_1,\varphi_0\rangle\zeta_1+\langle\chi_0,\varphi_0\rangle\zeta_0\Big)
    =\psi_0\otimes\tilde{\zeta}_0,
\end{array}\end{equation}
where $\{\tilde\eta_x\}$ and $\{\tilde\zeta_x\}$ are two bases in $\KK$ such that
\begin{equation}\nonumber
    \tilde\zeta_1=\frac{\tilde\eta_1+\tilde\eta_0}{\sqrt2}, \qquad
    \tilde\zeta_0=\frac{\tilde\eta_1-\tilde\eta_0}{\sqrt2},
\end{equation}
and so
\begin{equation}\nonumber
    T_\alpha(\ketbra{\varphi_x}{\varphi_x})
    = \tr_\KK\Big( U \,(\ketbra{\varphi_x}{\varphi_x}\otimes\frac{\idty}{2})\, U^*
    \,(\idty\otimes\ketbra{\mu_\alpha}{\mu_\alpha}) \Big)
    = \frac{1}{2} \Big\{ \ketbra{\psi_1}{\psi_1}\cdot|\langle\tilde\eta_x,\mu_\alpha\rangle|^2
    + \ketbra{\psi_0}{\psi_0}\cdot|\langle\tilde\zeta_x,\mu_\alpha\rangle|^2  \Big\},
\end{equation}
where at most one of the coefficients $|\langle\tilde\eta_1,\mu_\alpha\rangle|^2$,
$|\langle\tilde\eta_0,\mu_\alpha\rangle|^2$, $|\langle\tilde\zeta_1,\mu_\alpha\rangle|^2$,
$|\langle\tilde\zeta_0,\mu_\alpha\rangle|^2$ can vanish so that $T_\alpha(\ketbra{\varphi_1}{\varphi_1})$
and $T_\alpha(\ketbra{\varphi_0}{\varphi_0})$ are never orthogonal and $T$ is not `S'.

\section{Analysis with LOCC operations}\label{sec:locc}

Our correction scheme was motivated by situations where the  
measurement on the environment and subsequent recovery are 
realized by physical processes, possibly to be run repeatedly, or 
even in continuous time. Therefore we did not encorporate a 
possible dependence of the whole process on information previously 
gained on the system. On the other hand, the `lost and found' 
scenario might well be extended to contain such dependences. The 
most general case would then maybe begin with a quantum 
measurement on the system, to ascertain the losses. Based on that 
information a classical call would be made to the Lost and Found 
Office in the environment. An employee would then make a quantum 
measurement on the shelves in the office, maybe ask a question for 
more classical information about the system, and so on. All this 
amounts to a cooperation between the system manager and the Lost 
and Found employee, in the framework of an LOCC (`Local quantum 
operations and classical communication'') protocol \cite{locc}. 
Developing the corrigibility criteria for such protocols is beyond 
the present work. Therefore we only give one example showing that 
the extended framework does make a difference in some situations. 
 
We consider Example~8 of Section \ref{mixedenv}, where classical 
information encoded on the first qubit can never be retrieved 
after the interaction by our correction scheme. The simple LOCC 
scheme, which will achieve correction in this case is to start 
with a measurement on the `system' qubit followed by a suitable 
measurement on the `environment' qubit, chosen according to the 
first result. 

Fix the encoding basis $B=\{\varphi_x\}$ in $\HH$, the 
corresponding bases $\{\tilde\eta_x\}$ and $\{\tilde\zeta_x\}$ are 
fixed in $\KK$, too, and the initial state of the bipartite system 
is one of the $\ketbra{\varphi_x}{\varphi_x}\otimes\idty/2$, 
$x=1,0$. After the interaction $U$ we measure on the first qubit 
$F_\alpha = \ketbra{\psi_\alpha}{\psi_\alpha}\in\BB(\HH)$ so that, 
according to the result $\alpha$ and the classical information 
$x$, the subsequent (non normalized) state of the second qubit is 
\begin{equation}\nonumber 
    \tilde\rho_\alpha(x)
    = \tr_\HH\Big( U \,(\ketbra{\varphi_x}{\varphi_x}\otimes\frac{\idty}{2})\, U^*
       \,(\ketbra{\psi_\alpha}{\psi_\alpha}\otimes\idty) \Big)
    = \frac{1}{2} \ketbra{\tilde\xi_x^{(\alpha)}}{\tilde\xi_x^{(\alpha)}}, \qquad
       \mbox{where } \tilde\xi_x^{(\alpha)} 
    = \begin{cases} \tilde\eta_x, & \alpha=1, \\ \tilde\zeta_x,
    & \alpha=0. \end{cases}
\end{equation}
Therefore $\tilde\rho_\alpha(1)$ and $\tilde\rho_\alpha(0)$ are orthogonal for every $\alpha$ and
classical information can be recovered by a measurement of $\{\tilde\xi_x^{(\alpha)}\}_x$.

\par\vskip14pt \noindent\textbf{Acknowledgements}. M.G.\ 
gratefully acknowledges support from the Alexander von Humboldt 
Foundation. Our work was also supported by the European Union 
project EQUIP (contract IST-1999-11053) and the DFG (Bonn).


\begin{thebibliography}{ccccc} 
\bibitem{Davies}E.~B.~Davies: 
 Quantum theory of open system 
 (Academic Press, London, 1976) 

\bibitem{Nielsen}M.A.~Nielsen and I.L.~Chuang: Quantum computation 
and quantum information (Cambridge UP, Cambridge 2000) 

\bibitem{QECC}M. Keyl, R.F. Werner:  How to correct small quantum 
errors. 
 To appear in: A. Buchleitner and K. Hornberger (eds.), Coherent Evolution in Noisy Environment,
 Lecture Notes in Physics, Springer (2002); preprint quant-ph/0206086; 
 
\bibitem{Alber}G.~Alber, Th.~Beth, Ch.~Charnes, A.~Delgado, 
M.~Grassl, M.~Mussinger: Stabilizing distinguishable qubits 
against spontaneous decay by detected-jump correcting quantum 
codes. {\it Phys.\ Rev.\ Lett.} \textbf{86} No.~19, 4402-4405 
(2001) 

\bibitem{Plenio}M.~B.~Plenio, V.~Vedral, P.~L.~Knight:
 Quantum error correction in the presence of
spontaneous emission. {\it Phys.\ Rev.\ A} \textbf{55}, 67-71 (1997)

\bibitem{Landau}L.~J.~Landau, R.~F.~Streater: On Birkhoff's 
theorem for doubly stochastic completely positive maps of matrix 
algebras. {\it J.~Linear Alg.~Appl.} \textbf{193}, 107-127 (1993) 

\bibitem{Kummer}H. Maassen, B. K\"ummerer: The essentially 
commutative dilations of dynamical semigroups on $M_n$. 
 {\it Commun. Math. Phys.} {\bf 109}(1987)1-22; 
 see especially the remark after Proposition 2.2.1.

\bibitem{telepo}R.F. Werner:  All Teleportation and Dense Coding Schemes.
{\it J. Phys. A}\/ {\bf 35} (2001) 7081-7094 or quant-ph/0003070

\bibitem{BarRac}A.~O.~Barut, R.~Raczka: Theory of Group 
Representations and Applications (World Scientific Publishing, 
Singapore, 1986) 

\bibitem{holevoW} A.S. Holevo, R.F. Werner:
 Counterexample to an additivity conjecture for output purity of quantum channels.
 To appear in {\it J.Math.Phys.} (2002), see  quant-ph/0203003

\bibitem{unot}V. Buzek, M. Hillery, R.F. Werner: 
 Optimal Manipulations with Qubits: Universal NOT Gate.
{\it Phys. Rev. A} {\bf 60}, R2626 (1999) or quant-ph/9901053


\bibitem{WSHV}J.~Walgate, A.~J.~Short, L.~Hardy, V.~Vedral:
 Local distinguishability of multipartite orthogonal quantum states. 
 {\it Phys.\ Rev.\ Lett.} \textbf{85}, 4972 (2000)

\bibitem{locc}C. Bennett, H. Bernstein, S. Popescu, B. Schumacher: 
Concentrating partial entanglement by local operations, {\it Phys 
Rev A \bf53} (1996) 2046, or quant-ph/9511030. 

 \end{thebibliography}
 \end{document}